\documentclass[aps, prd, superscriptaddress, nofootinbib]{revtex4}

\usepackage{graphicx}
\usepackage{dcolumn}
\usepackage{amssymb}
\usepackage{hyperref}
\usepackage[all]{xy}
\usepackage{epsfig}
\usepackage{graphicx}
\usepackage{amsfonts,amsmath}
\usepackage{subfigure}

\usepackage[dvips]{color} 

\begin{document}

\newcommand{\nab}[1]{\textcolor[rgb]{1,0,0}{#1}}

\newcommand{\conj}{^{*}}
\newcommand{\half}{\frac{1}{2}}
\newcommand{\vect}[1]{\mathbf{#1}}

\newcommand\beq{\begin{equation}}
\newcommand\eeq{\end{equation}}
\newcommand\bea{\begin{eqnarray}}
\newcommand\eea{\end{eqnarray}}
\newcommand\be{\begin{equation}}
\newcommand\ee{\end{equation}}
\newcommand\del{\partial}
\newcommand\cM{{{\cal{M}}}}
\newcommand\ba{{\mathbf{a}}}
\newcommand\bb{{\mathbf{b}}}
\newcommand\bff{{\mathbf{f}}}
\newcommand\bx{{\mathbf{x}}}
\newcommand\bxd{\dot{\mathbf{x}}}
\newcommand\bX{{\mathbf{X}}}
\newcommand\bXA{{\mathbf{X}^A}}
\newcommand\bXB{{\mathbf{X^B}}}
\newcommand\bXd{\dot{\mathbf{X}}}
\newcommand\bzero{{\mathbf{0}}}
\newcommand\pa{\partial}
\newcommand\nn{\nonumber}
\newcommand\cd{\cdot}
\newcommand\al{\alpha}
\newcommand\ga{\gamma}
\newcommand\gau{\gamma_u}
\newcommand\de{\delta}
\newcommand\ep{\epsilon}
\renewcommand\th{\theta}
\newcommand\si{\sigma}
\newcommand\ta{\tau}
\newcommand\lra{\leftrightarrow}
\newcommand{\gsim}{\raise.3ex\hbox{$>$\kern-.75em\lower1ex\hbox{$\sim$}}}
\newcommand{\lsim}{\raise.3ex\hbox{$<$\kern-.75em\lower1ex\hbox{$\sim$}}}
\newcommand\eq{&\!\!\!=\!\!\!&}
\renewcommand\th{\theta}
\newcommand\rh{\rho}
\renewcommand\rhd{\dot\rho}
\def\d{{\rm d}}

\newcommand{\figname}[1]{\textit{#1}}

\setcounter{footnote}{0} \renewcommand{\thefootnote}{\arabic{footnote}}

\title{Strong coupling in Ho\v rava gravity.}

\newcommand{\addressNott}{School of Physics \& Astronomy, University of Nottingham, University Park, Nottingham, NG7 2RD, United Kingdom.}
\newcommand{\addressPS}{Laboratoire de Physique Theoretique,
  Universtite Paris Sud-11,91405 Orsay Cedex, France.}
\newcommand{\addressTour}{LMPT
 UFR Sciences et Techniques, Universite Francois Rabelais, Parc de Grandmont
 37200 Tours, France.}

\author{Christos Charmousis}
\email{Christos.Charmousis@th.u-psud.fr}
\affiliation{\addressPS}
\affiliation{\addressTour}

\author{Gustavo Niz}
\email{gustavo.niz@nottingham.ac.uk}
\affiliation{\addressNott}

\author{Antonio Padilla}
\email{antonio.padilla@nottingham.ac.uk}
\affiliation{\addressNott}

\author{Paul M. Saffin}
\email{paul.saffin@nottingham.ac.uk}
\affiliation{\addressNott}

\date{\today}

\begin{abstract}
By studying perturbations about the vacuum, we show that Ho\v rava gravity suffers from two different strong coupling problems, extending all the way into the deep infra-red. The first of these is associated with the principle of detailed balance and explains why solutions to General Relativity are typically not recovered in models that preserve this structure. The second of these occurs even without detailed balance and is associated with the breaking of diffeomorphism invariance, required for anisotropic scaling in the UV.  Since there is a reduced symmetry group there are additional degrees of freedom, which need not decouple in the infra-red. Indeed, we use the Stuckelberg trick to show that  one of these extra modes become strongly coupled as the parameters approach their desired infra-red fixed point. Whilst we can evade the first strong coupling problem by breaking detailed balance, we cannot avoid the second, whatever the form of the potential. Therefore the original Ho\v rava model, and its "phenomenologically viable" extensions {\it do not have a perturbative General Relativity limit at any scale}.  Experiments which confirm the perturbative gravitational wave prediction of General Relativity, such as the cumulative shift of the periastron time of binary pulsars, will presumably rule out the theory.

\end{abstract}

\maketitle

\section{Introduction} \label{sec:intro}
Ho\v rava has recently proposed an interesting toy model of quantum
gravity~\cite{Horava:2008ih,Horava:2009uw, Horava:2009if}, generating
a whole slew of publications that examine various aspects of the
theory (see, for example~\cite{kofinas, everything, everything2,
  Lu:2009em, Kehagias:2009is, Sotiriou:2009gy}). At short distances
the theory describes interacting nonrelativistic gravitons, and is
argued to be power counting renormalisable in $3+1$
dimensions. Relativistic physics is supposed to emerge in the
infra-red via relevant deformations, such that General Relativity is
recovered at large distances. Since Lorentz symmetry is manifestly
broken in this theory, there are, in general,  a huge number of
possible relevant deformations one could include. To restrict the
number of possible parameters in the model,  Ho\v rava made use of the
principle of "detailed balance", as developed in studies of
non-equilibrium critical phenomena and quantum critical systems.
Whilst this organising principle is elegant,  it would appear to be an
obstacle to recovering GR in the infra-red. This was first illustrated
in a study of static spherically symmetric solutions that did not
recover the Schwarzschild geometry at large distances, unless
detailed balance was broken~\cite{Lu:2009em,Kehagias:2009is}. This has
led to so called "phenomenologically viable" extensions of the model
that break detailed balance explicitly~\cite{Sotiriou:2009gy}. 

In this paper we will show that Ho\v rava gravity suffers from strong coupling problems, with and without detailed balance, and is therefore unable to reproduce General Relativity in the infra-red. We consider the perturbative theory about the vacuum, yielding two important results. The first considers the role of detailed balance in these models. As the breaking terms go zero, we find that the linearised  gravitational Hamiltonian constraint vanishes {\it off}-shell. This means that linearised theory breaks down in this limit, just as it does for the Chern-Simons limit of Gauss-Bonnet gravity~\cite{GB} (for a review on these gravity theories see \cite{zanelli}, \cite{cc}). By comparing our equations to their counterparts in General Relativity, we can see that the "emergent" Planck length actually diverges in the limit of detailed balance, in contrast to the original claims~\cite{Horava:2009uw}.  This strong coupling behaviour means that the theory with detailed balance does not have a perturbative infra-red limit of any sort,  explaining the results of~\cite{Lu:2009em}. Indeed, from the point of view of spherically symmetric solutions one sees that the putative higher order terms in the IR are  just as important as the "lower" order terms. In summary, with detailed balance, we can never hope to recover  GR in the infra-red for the following reason:  General Relativity admits an effective linearised description beyond the Schwarzschild radius of a source,  but in Ho\v rava gravity with detailed balance, strong coupling prevents an effective linearised description on {\it any} scale.

Our second result also applies to those models that have been dubbed "phenomenologically viable", and break detailed balance explicitly. In some sense it is clear that breaking detailed balance cannot possibly be enough to recover GR in these models. The point is that General Relativity contains full diffeomorphism invariance, so that the theory has just two propagating degrees of freedom. Because Lorentz symmetry is necessarily broken in the UV, Ho\v rava gravity contains a reduced set of diffeomorphisms, and must therefore contain more propagating degrees of freedom. If GR is to be recovered in the infra-red, these extra degrees of freedom should decouple from the system. This is not what happens. By restoring the full set of diffeomorphisms using the Stuckelberg trick, we are able to show that one of the additional degrees of freedom actually becomes strongly coupled as the parameters in the theory flow towards their desired infra-red fixed points. The scenario is highly reminiscent of Pauli-Fierz massive gravity~\cite{PF} in which the longitudinal scalar becomes strongly coupled as $m \to 0$~\cite{PFstrong}, leading to the famous vDVZ discontinuity~\cite{vdvz}. This result is  independent of how one chooses to break detailed balance.

%


\section{Anisotropic scaling and Ho\v rava gravity} \label{sec:horavaGravity}

We begin by reviewing the basic ideas behind Ho\v rava gravity in scalar field theory, using Lifshitz's model for a scalar field that explicitly breaks Lorentz invariance~\cite{lifshitz} (see also~\cite{Anselmi:2007ri,Visser:2009fg}). This provides a different way to regulate the UV divergences of loop diagrams, avoiding violations of unitarity associated with Pauli-Villars and higher derivative Lorentz invariant regulators, and without the need to introduce ficticious non-integer dimensions as in dimensional regularisation. The hope then is that while Lorentz symmetry is explicity broken at high energy scales, it may be recovered in the IR regime at low energies. Consider, for example, the action~\cite{Visser:2009fg}
\be
S_{free}=\int dt\;d^Dx\left[\half\dot\phi^2-\half\phi(-\nabla^2)^z\phi)\right], \label{Sfree}
\ee
This describes a free field fixed point with anisotropic scaling between space and time,
\be
x^i \to l x^i, \qquad t \to l^z t,
\ee
characterised by the 'dynamical critical exponent' $z$, so that the
scaling dimensions are $[x]=-1$ and $[t]=-z$. The action (\ref{Sfree})
leads to internal propagators in the UV of the form
\bea
G(\omega,\underline k)\rightarrow |\underline k|^{-2z}.
\eea
For large enough $z$ one sees that the fall-off of the propagator is fast enough to render Feynman diagrams convergent. In fact, the superficial degree of divergence, $\delta$, satisfies
\bea
\delta\leq(D-z)L,
\eea
for $L$ loops~\cite{Visser:2009fg}. As it stands, this model is not acceptable because it has no Lorentz symmetry in the IR. This may be remedied by including a relevant operator of the form
\bea
S_{rel}&=&\int dt\;d^Dx\left[-\half c_{(\phi)}^2\del^i\phi\del_i\phi\right],
\eea
leading to a model that flows to a theory with Lorentz symmetry emergent at low energies, with a light-cone defined by the parameter $c_{(\phi)}$. It is interesting to note that if we have a number of matter fields, they can each have their own Lorentz symmetry. This is not something that is observed experimentally and  leads to a fine tuning of the model. There are further issues that appear once Lorentz symmetry is broken, such as the possibility of a black-hole perpetuum mobile machine~\cite{Jacobson:2008yc,Dubovsky:2006vk}. Furthermore, although the action $S_{free}+S_{rel}$ breaks Lorentz invariance in the UV, it does not introduce extra degrees of freedom in the infra-red as the emergent symmetry is not dynamical. In General Relativity, however, diffeomorphism invariance is a dynamical symmetry, so breaking it in the UV could alter the number of degrees of freedom that propagate in the infra-red. This leads directly to the second strong coupling problem alluded to earlier.

With Lorentz symmetry no longer being used as a guiding principle,
there is a great proliferation in the number of terms that may appear
in the action. To ameliorate this, Ho\v rava proposed an organising
principle based on {\it detailed balance}~\cite{Horava:2009uw}, which
also allows one to put forward a {\it quantum inheritance} principle
such that the theory in $D+1$ dimensions acquires the renormalisation
properties of the $D$-dimensional theory~\cite{Horava:2008jf}. Detailed
balance is a statement that the potential of a $D+1$-dimensional
theory is obtained from a $D$-dimensional "superpotential" by
functional differentiation. For example, the scalar field action is
given by
\be
S=\int dt\;d^Dx\left[\half\dot\phi^2-\half\left(\frac{\delta W}{\delta\phi}\right)^2\right],
\ee
with a superpotential
\be
W[\phi]=\int d^Dx\left[\frac{1}{2}\del_i\phi\del^i\phi+\half m\phi^2\right].
\ee
In this case, one obtains a $z=2$ theory in the UV, with a Lorentz invariant Klein Gordon theory in the IR, with an "emergent" speed of light $c^2_{(\phi)}=2m$. 
In the first of our results, we shall  find that the gravity model constructed using detailed balance does not have a well defined perturbation limit about its vacuum. However, we ought to note that this is not an artefact of detailed balance in general. For example, the scalar model above has well defined wave solutions in the vacuum
\be
\phi(t,\underline x)=e^{i(\omega t+\underline k.\underline x)},\qquad
  \omega=\pm(|\underline k|^2+m).
\ee
Since the theory is linear, the perturbations are also waves. The situation in gravity is rather more subtle owing to the fact that there are constraints, specifically the Hamiltonian constraint, as we shall see in the next section. The constraint equations lead to strong coupling unless detailed balance is broken.

The gravitational theory based on the violation of Lorentz symmetry has been clearly presented in Ho\v rava's paper~\cite{Horava:2009uw} and we refer the reader to that work for more detail.  One of Ho\v rava's key assumptions is the explicit breaking of  full four dimensional diffeomorphism  invariance to a subgroup that preserves a foliation structure of space-like slices. This enables him to make use of anisotropic scaling in the UV as in the Lifshitz model we have just discussed. Following from the ADM decomposition of the metric, and the Einstein equations~\cite{Arnowitt:1960es}, the fundamental objects of interest are the fields $N(t,\underline x)$, $N_i(t,\underline x)$, $g_{ij}(t,\underline x)$, corresponding to the lapse, shift and spatial metric of the ADM decomposition,
\be
ds^2 =\hat g_{\mu\nu} dx^{\mu}dx^\nu=-N^2 c^2dt^2+g_{ij}(dx^i+N^i dt)(dx^j+N^j dt).
\ee
Under the new, restricted, set of diffeomorphisms
\be
x^i \to x^i-\zeta^i(t, \underline x), \qquad t \to t -f(t)
\ee
the fields transform as follows
\bea
\delta g_{ij}& \to & \delta g_{ij}+2 \nabla_{(i} \zeta_{j)}+f\dot g_{ij}, \label{shiftHh}\\
\delta N_i&\to & \delta N_i+ \del_i(\zeta^jN_j)-2\zeta^j\nabla_{[i}N_{j]}+\dot\zeta^jg_{ij}+\dot fN_i+f\dot N_i,  \label{shiftHni}\\
\delta N&\to &\delta N+\zeta^j\del_jN+\dot fN+f\dot N.  \label{shiftHn}
\eea
where indices are raised/lowered using $g_{ij}$, and $\nabla_i$ is the covariant derivative on the space-like slices.

The transformation laws represent an important deviation from standard General Relativity, where full $4D$ diffeomorphism invariance is present.  Indeed, note that the last of these transformations shows that if $N$ is restricted to be "projectable"~\cite{Horava:2009uw}, i.e. $N=N(t)$, then this condition is maintained under the restricted diffeomorphism group. Projectable solutions to Ho\v rava's theory cannot, therefore,  be transformed into non-projectable solutions, in contrast to General Relativity.  This explicitly illustrates the fact that solutions to Ho\v rava gravity cannot be specified using the  $4D$ metric alone--one must always specify the foliation. Furthermore, although one is free to impose projectability at the level of solutions in Ho\v rava gravity, doing so prevents us from finding the full set of solutions. Again, this is not the case in GR where one can always use the full set of diffeomorphisms to render any solution locally projectable.

 In this paper, we shall consider the general case, as Ho\v  rava does,  where $N$ is a function of both  $x^i$ and $t$. We note that imposing projectability at the level of theory, as advocated in~\cite{Sotiriou:2009gy}, alters the theory explicitly, since the equations of motion for the lapse can only then be expressed as integrals over space. Such a modification of Ho\v rava gravity would appear to be inherently non-local, so we will not consider it here.

The action for Ho\v rava gravity is made up of a kinetic term, and a potential term satisfying "detailed balance",
\be
S_{H}=\int dt d^3 x  \sqrt{g} N (T-V) \label{action}.
\ee
The kinetic term is constructed out of the extrinsic curvature of the foliations, as this is covariant under the remnant diffeomorphism symmetry,
\bea
K_{ij}&=&\frac{1}{2N}(\dot g_{ij}-\nabla_iN_j-\nabla_jN_i).
\eea
Requiring the kinetic term to be at most quadratic in $K$ yields
\bea
T&=&\frac{2}{\kappa^2}(K_{ij}K^{ij}-\lambda K^2)=\frac{2}{\kappa^2}K_{ij}G^{ijkl}K_{kl}
\eea
where we have introduced the the de Witt metric,
\be
G^{ijkl}=\half(g^{ik}g^{jl}+g^{il}g^{jk})-\lambda g^{ij}g^{kl},
\ee
whose inverse is given by
\be
G_{ijkl}=\half(g_{ik}g_{jl}+g_{il}g_{jk})-\tilde\lambda g_{ij}g_{kl},\qquad\qquad
\tilde\lambda=\frac{\lambda}{3\lambda-1}.
\ee
The dimensionless parameter $\lambda$ is taken to run with scale. In order to have any hope of recovering General Relativity in the IR, one must assume that $\lambda=1$ corresponds to the infra-red fixed point. The potential term is constructed out of the spatial metric and its derivatives. Inspired by methods used in quantum critical systems and non-equilibrium critical phenomena, Ho\v rava restricts the large class of possible potentials using the principle of detailed balance outlined above. This requires that the potential takes the form
\bea
V&=&\frac{\kappa^2}{8}\frac{1}{\sqrt{g}}\frac{\delta W}{\delta g_{ij}}G_{ijkl}\frac{1}{\sqrt{g}}\frac{\delta W}{\delta g_{kl}}\\
    &=&\frac{\kappa^2}{8}E^{ij}G_{ijkl}E^{kl}.
\eea
Note that by constructing $E^{ij}$ as a functional derivative it automatically becomes transverse from within the foliation slices, $\nabla_iE^{ij}=0$.
We can derive the field equations by varying the action (\ref{action}) with respect to each of the fields~\cite{kofinas},
\bea
\frac{1}{\sqrt{g}}\frac{ \delta S_{H}}{\delta N} &=& -(T+V), \\
\frac{1}{\sqrt{g}}\frac{ \delta S_{H}}{\delta N_i} &=& \frac{4}{\kappa^2} \nabla_i \pi^{ij}, \\
\frac{1}{\sqrt{g}}\frac{ \delta S_{H}}{\delta g_{ij}} &=&  -\frac{2}{\kappa^2}\left[ \dot \pi^{ij}+NK\pi^{ij} +2\nabla_k(\pi^{k(i} N^{j)})-N_k \nabla^k \pi^{ij} +2N K^{ki}\pi^j_k\right] +\frac{1}{2} N(T-V)g^{ij}\nonumber\\
&& \qquad -\frac{\kappa^2}{4}\left[ \Delta(N \chi^{ij})+NE^i_k \chi^{jk}\right],
\eea
where
\be
\pi^{ij}=K^{ij}-\lambda Kg^{ij}, \qquad
\chi^{ij} = E^{ij}-\tilde \lambda Eg^{ij} ,
\ee
and the operator $\Delta$ is defined as\footnote{To illustrate what we mean by this definition, note that we can define the Lichnerowicz operator in a similar way,
$-\frac{1}{2}\Delta_L h_{ij}=\lim_{\epsilon \to 0} \frac{1}{\epsilon} \left( R_{ij}[g+\epsilon h]-R_{ij}[g]\right)$.}
\be
\Delta h^{ij}=\lim_{\epsilon \to 0} \frac{1}{\epsilon} \left( E^{ij}[g+\epsilon h]-E^{ij}[g]\right).
\ee
Having constructed the gravitational theory following the same principles as those for the scalar field,  it remains to pick the superpotential $W[g_{ij}]$.  In $3+1$ dimensions, this must be chosen such that we have anisotropic scaling with a dynamical critical exponent $z \geq 3$, in order that the theory be power counting renormalisable. This follows from the fact that in D+1-dimensions, the scaling dimension of $\kappa$ is given by
\be
[\kappa]=\frac{z-D}{2}.
\ee
Fully relativistic theories such as general relativity must always have $z=1$. In the next section we will focus on the case of $z=3$, so that $\kappa$ is dimensionless in $3+1$ dimensions.

\section{z=3 Ho\v rava gravity with and without detailed balance} \label{sec:strongHoravaGravity}
Given the guiding principle of detailed balance, the unique $z=3$
theory in $3+1$ dimensions, with additional relevant deformations in
the IR may be obtained from the following superpotential~\cite{Horava:2009uw},
\be
W[g_{ij}]=\frac{1}{w^2}\int  \omega_3(\Gamma)+\mu \int d^3x \sqrt{g}(R-2\Lambda_W).
\ee
The $z=3$ contribution comes from  gravitational Chern-Simons action in 3-dimensions, where
\be
\omega_3(\Gamma)={\textrm Tr} \left(\Gamma \wedge d \Gamma +\frac{2}{3} \Gamma \wedge \Gamma  \wedge \Gamma \right).
\ee
Again, all the couplings are taken to run with scale, with scaling dimensions  $[w]=0,~ [\mu]=1, ~[\Lambda_W]=2$. Variation of this action yields
\be
E^{ij}=\frac{1}{w^2}C^{ij}-\frac{\mu}{2}\left(G^{ij}+\Lambda_W g^{ij}\right),
\ee
where $G^{ij}$ is the Einstein tensor on the spatial slices, and $C^{ij}$ is the Cotton tensor
\be
C^{ij}=\epsilon^{kl(i} \nabla_k R^{j)}_l.
\ee
Ho\v rava originally argued that this theory flowed from $\lambda=1/3$
in the UV, to $\lambda=1$ in the IR, thereby recovering General
Relativity at low energies, with an emergent speed of light, $c$,
Newton's constant, $G_N$, and cosmological constant, $\Lambda$, given by
\be
c=\frac{\kappa^2 \mu}{4}\sqrt{\frac{\Lambda_W}{1-3\lambda}}, \qquad G_N=\frac{\kappa^2}{32\pi c}, \qquad \Lambda=\frac{3}{2}\Lambda_W.
\ee
However, a study of spherically symmetric solutions in this theory~\cite{Lu:2009em} seems to indicate that this is not the case. One has to break detailed balance in order to recover the corresponding solutions in General Relativity. We will now show  that this is because detailed balance leads to strong coupling on all scales, so that one cannot consistently truncate the higher derivative operators in the infra-red. To elucidate the specific role played by detailed balance let us break it explicitly. Clearly there are a number of ways in which one can do this. A  set of relevant breaking terms was proposed in~\cite{Sotiriou:2009gy}, although we note here that their list did not include terms like $\int dt d^3x \sqrt{g}N C^{ij}R_{ij}$, which seem perfectly reasonable at first glance. For simplicity, we will perform a minimal breaking of detailed balance by adding a term to the action of the form
\be
S_{br}=-\frac{\kappa^2}{8} \left(\frac{\epsilon}{1-3\lambda}\right) \int dt d^3 x \sqrt{g}N  (R-3\beta)
\ee
where, from the point of view of the $z=3$ theory at short distances, the new parameters have scaling dimension $[\epsilon]=4$, $[\beta]=2$. We will also include a generic matter contribution, $S_m$, so that the full action is now given by
\be
S=S_H+S_{br}+S_m.
\ee
Of course, it is not exactly clear how we should couple matter in this theory, as we no longer have the guiding hand of Lorentz invariance to assist us. We will not worry about those issues here, merely assuming that it can be done in some consistent way, so that the matter fields act as sources in our equations of motion.  The field equations now take the form
\bea
\frac{1}{\sqrt{g}}\frac{ \delta S_{H}}{\delta N} -\frac{\kappa^2}{8} \left(\frac{\epsilon}{1-3\lambda}\right) (R-3\beta)&=& - \frac{1}{\sqrt{g}}\frac{ \delta S_{m}}{\delta N}\label{N}=\rho,\\
\frac{1}{\sqrt{g}}\frac{ \delta S_{H}}{\delta N_i} &=&-\frac{1}{\sqrt{g}}\frac{ \delta S_{m}}{\delta N_i}= v^i \label{Ni}, \\
\frac{1}{\sqrt{g}}\frac{ \delta S_{H}}{\delta g_{ij}} -\frac{\kappa^2}{8} \left(\frac{\epsilon}{1-3\lambda}\right) \left[\nabla^i\nabla^j-(g^{ij}\nabla^2+G^{ij}+\frac{3\beta}{2}g^{ij})\right]N&=& -\frac{1}{\sqrt{g}}\frac{ \delta S_{m}}{\delta g_{ij}}=\tau^{ij} \label{gij}.
\eea
The energy-momentum fields of the matter contribution $(\rho, v^i, \tau^{ij})$ satisfy the following conservation laws
\bea
\int d^3x\sqrt{g} \left[\dot g_{ij} \tau^{ij}-N\frac{(\rho \sqrt{g})\dot{}}{\sqrt{g}}-N_i\frac{(v^i \sqrt{g})\dot{}}{\sqrt{g}}\right]&=&0 \label{con1}, \\
2\nabla^i \tau_{ij}-\rho\del_jN+\frac{(v^i \sqrt{g})\dot{}}{\sqrt{g}}+N_j\nabla_iv^i+2v^i\nabla_{[i} N_{j]} &=&0 \label{con2}.
\eea
These deviate slightly from the usual conservation of energy-momentum, $\hat \nabla_\mu T^{\mu\nu}=0$,   because we only have the reduced set of diffeomorphisms outlined in the previous section.

We now wish to define vacua\footnote{We will denote vacuum expectation
 values for all fields with a ``bar''.}  in this theory, in the absence
of these matter fields. Owing to the fact that we have a reduced set
of diffeomorphisms, it is not enough to impose, say, maximal symmetry
in $3+1$ dimensions. We must also define the foliation. To this end we
note that the momentum conjugate to $g_{ij}$ is given by
$p^{ij}=\sqrt{g} \pi^{ij}$, and require it to vanish on the vacuum, so
that $\bar  K_{ij}=0$. Further, we choose the gauge $\bar N_i=0$, and
require that the spatial metric, $\bar g_{ij}$ is a homogeneous
Einstein space
\be
\bar g_{ij}dx^i dx^j=\frac{dr^2}{1-\frac{\gamma}{2} r^2}+r^2 d\Omega_2
\ee
with constant Ricci curvature $\bar R_{ij}=\gamma g_{ij}$. In geometric terms we are asking for our 3 dimensional foliation to be maximally symmetric and furthermore that the foliation be trivially embedded (totally geodesic). Given that 3-space is conformally flat, $\bar E^{ij}=q \bar g^{ij}$ where $q=\frac{\mu}{4}(\gamma-2\Lambda_W)$. The $N_i$ equation (\ref{Ni}) is satisfied automatically. The $N$ equation (\ref{N}), which is essentially the Hamiltonian constraint, yields $q^2=\epsilon(\beta-\gamma)$, and so
\be
\gamma=2\Lambda_W-\frac{8}{\mu^2}\left(\epsilon \pm \sqrt{\epsilon^2+\frac{\epsilon \mu^2}{4}(\beta-2\Lambda_W)}\right).
\ee
It remains to impose the $g_{ij}$ equation (\ref{gij}), which constrains the background lapse function $\bar N$. The quantity $\Delta N \bar g^{ij}$ is easily derived by making use of the transformation laws for $G^{ij}$ and $C^{ij}$ under conformal transformations. We find that
\be
-\frac{\kappa^2}{16}\left(\frac{\mu q+2\epsilon}{1-3\lambda}\right) \left[\nabla^i\nabla^j-g^{ij}(\nabla^2+\gamma)\right]\bar N=0. \label{eomN}
\ee
For detailed balance, we have $\epsilon=q=0$, and so $\bar N$ is unconstrained. This is consistent with the findings of~\cite{Lu:2009em}. Away from detailed balance, we find that $\bar N =\sqrt{1-\gamma r^2/2}$, so that the full $3+1$ dimensional metric corresponds to a maximally symmetric spacetime with curvature $\gamma/2$,  written in global coordinates.

Let us now reintroduce the matter fields, and consider perturbations about the vacuum
\be
\delta N=n(t, \underline x), \qquad \delta N_i=n_i(t, \underline x), \qquad \delta g_{ij}=h_{ij}(t, \underline x).
\ee
It is convenient to introduce ${\cal E}^{ij}=E^{ij}-qg^{ij}$, as this vanishes on the background. The unbroken potential now takes the form
\be
V=\frac{\kappa^2}{8}\left[{\cal E}^{ij}G_{ijkl}{\cal E}^{kl}+2q(1-3 \tilde \lambda){\cal E}+3q^2(1-3\tilde \lambda)\right],
\ee
and the Hamiltonian constraint (\ref{N}) may be written
\be
-\frac{2}{\kappa^2}K_{ij}G^{ijkl}K_{kl}-\frac{\kappa^2}{8}\left[{\cal E}^{ij}G_{ijkl}{\cal E}^{kl}+\frac{1}{2}\left(\frac{1}{1-3\lambda}\right)(\mu q+2\epsilon) (R-3\gamma)\right]=\rho
\ee
where we have used the fact that $q^2=\epsilon(\beta-\gamma)$ and ${\cal E}=\frac{\mu}{4}(R-3\gamma)$. Perturbing this equation to linear order is now easy, since the first two terms are already second order owing to the fact that both $K_{ij}$ and ${\cal E}^{ij}$ vanish on the background. Lumping all higher order corrections alongside the matter field, the  Hamiltonian constraint  gives
\be
-\frac{\kappa^2}{16}\left(\frac{\mu q+2\epsilon}{1-3\lambda}\right) \delta R=\rho \label{hamcon}+\textrm{non-linear corrections}.
\ee
For detailed balance ($\epsilon=q=0$), we immediately see that linearised perturbation theory is not well defined in the presence of matter. Higher order terms always dominate,  and one loses predictive power. This is characteristic of strong coupling, and is reminiscent of the Chern-Simons limit in Gauss Bonnet gravity~\cite{GB}.  Perturbation theory around the vacuum is strongly coupled on all scales, even in the deep infra-red.  We have included a matter component to render this explicit, although it ought to be clear that vacuum fluctuations will also be strongly coupled since generically one does not expect all non-linear corrections to vanish identically.   Of course, one might hope to alleviate this strong coupling problem by perturbing about a different background. However, on temporal/spatial scales that are small compared to the scale set by  the background extrinsic curvature/spatial curvature, our vacuum solution would represent a good approximation for the background, and one would immediately lose predictability. For example, cosmological perturbations about an FRW background would become strongly coupled on subhorizon scales.

Of course, one can avoid this problem by moving away from detailed balance.  Indeed, it is instructive to compare equation (\ref{hamcon}) with the corresponding equation in General Relativity
\be
-\frac{c}{16\pi G_N} \delta R=\rho \label{grhamcon}+\textrm{non-linear corrections}.
\ee
This suggests that if General Relativity is indeed recovered in the infra-red, it does so with an emergent Newton constant $G_N=\kappa^2/32\pi c$ and an emergent speed of light
\be
c=\frac{\kappa^2}{4}\sqrt{\frac{\epsilon+\mu q/2}{1-3\lambda}}=\frac{\kappa^2}{4}\sqrt{\frac{\mp\left(\epsilon^2+\frac{\epsilon \mu^2}{4}(\beta-2\Lambda_W)\right)^{1/2}}{1-3\lambda}}. \label{c}
\ee
We immediately see that the upper branch of solutions is ruled out, as the emergent speed of light is imaginary. Even on the lower branch, as one approaches detailed balance $c\to 0$, and so $G_N \to \infty$, which  means the effective Planck length, $l_{pl}=\sqrt {\bar h G_N/c^3}$, diverges, as expected due to strong coupling on all scales. Away from detailed balance, strong coupling only kicks in below the emergent Planck length, and it is natural to ask if indeed General Relativity can be recovered in the infra-red, as is perhaps suggested by the form of equation (\ref{hamcon}).  To establish this properly we must also look at the linearised $N_i$ and $g_{ij}$ equations, and compare them with their GR counterparts. An entirely equivalent, but more convenient approach,  however, is to simply compute the effective action to quadratic order in the fields propagating on the background. We shall do this presently.

Let us rewrite the action as the emergent GR piece, plus corrections
\be
S=S_{GR}+S_{UV}+S_m,
\ee
where
\bea
S_{GR}&=&\frac{1}{16 \pi Gc} \int dt d^3x \sqrt{g}N\left[ K_{ij} K^{ij}-K^2-c^2(R-3\gamma)\right], \\
S_{UV}&=& \int dt d^3x \sqrt{g}N\left[ \frac{\kappa^2}{2}(1-\lambda)K^2-\frac{\kappa^2}{8}{\cal E}^{ij}G_{ijkl}{\cal E}^{kl}\right].
\eea
It is sufficient to compute $S_{UV}$ and $S_m$ to quadratic order. The latter is given by
\be
\delta_2 S_m= -\int dt d^3x \sqrt{\bar g}\left[n\rho+n_iv^i+h_{ij}\tau^{ij}\right].
\ee
Because $K_{ij}$ and ${\cal E}^{ij}$ vanish on the background, it is also straightforward to compute
\be
\delta_2 S_{UV}= \int dt d^3x \sqrt{\bar g}\bar N\left[ \frac{\kappa^2}{2}(1-\lambda) (\delta K)^2-\frac{\kappa^2}{8}\delta {\cal E}^{ij}\bar G_{ijkl}\delta {\cal E}^{kl}\right]
\ee
where
\be
\delta K=\frac{1}{2\bar N}\left[ \dot h-2 \nabla^i n_i \right], \qquad
\delta {\cal E}^{ij} = \frac{1}{w^2}\epsilon^{kl(i} \nabla_k \psi^{j)}_l-\frac{\mu}{2}\psi^{ij}
\ee
and
\be
\psi^{ij}= \delta\left(G^{ij}+\frac{\gamma}{2} g^{ij}\right)=-\frac{1}{2} \nabla^2 (h^{ij}-h\bar g^{ij})+\nabla^{(i} \nabla_{k} h^{j)k}-\frac{1}{2} \nabla^i\nabla^j h+\bar g^{ij} \nabla_k \nabla_l h^{kl}+\frac{\gamma}{2} h^{ij}
\ee
Assuming that $\lambda$ flows to $1$ in the infra-red, it would appear that
\be
\delta_2 S_{UV} \to -\frac{\kappa^2}{8}  \int dt d^3x \sqrt{\bar g}\bar N\delta {\cal E}^{ij}\bar G_{ijkl}\delta {\cal E}^{kl}.
\ee
This piece contains contributions that are higher order in the appropriate derivative operators, and can be ignored at low energies, compared with $\delta_2 S_{GR}$.   This would  suggest
 that provided we break detailed balance, we can indeed recover General Relativity at low energies. However, such a naive analysis clearly does not tell the full story. Recall that our original theory was invariant under a reduced set of diffeomorphisms.  This means the theory should contain more degrees of freedom than General Relativity. If our theory is to recover GR in the infra-red, where did the extra degrees of freedom go? One faces a similar scenario when studying Pauli-Fierz massive gravity theories~\cite{PF}.  A massive graviton has 5 propagating degrees of freedom whereas as a massless graviton has just two. As we take the graviton mass to zero in Pauli Fierz theory, the extra 3 degrees of freedom do not all disappear. In fact, it turns out that the longitudinal scalar mode becomes strongly coupled~\cite{PFstrong}, and is responsible for the famous vDVZ discontinuity~\cite{vdvz}.

The behaviour of the additional degrees of freedom in Pauli-Fierz
gravity is most clearly understood by artificially restoring the full
gauge invariance using the Stuckelberg trick~\cite{PFstrong}. This was
first introduced to study massive Abelian gauge theories, although we
shall apply it to the case in hand.  To begin with, note that under
the full set of diffeomorphisms present in General Relativity, $(t,
x^i) \to (t- f(t, \underline x), x^i-\zeta^{i}(t, \underline x))$, our ADM variables transform on the background as follows
\bea
n &\to& n+\zeta^k\nabla_k \bar N+\dot f \bar N +f \dot{\bar N} \label{shiftn},\\
n_i &\to& n_i+\dot \zeta^j \bar g_{ij}-\bar N^2c^2\del_i f \label{shiftni},\\
h_{ij} &\to& h_{ij}+2\nabla_{(i} \zeta_{j)} \label{shifth}.
\eea
We now introduce the Stuckelberg fields $\xi^i(t, \underline x)$, $\phi(t, \underline x)$, whose scaling dimensions are the same as $x$ and $t$ respectively. If we perform the following field redefinitions in the action
\bea
n &\to&   n+\xi^k\nabla_k \bar N+\dot \phi \bar N +\phi \dot{\bar N}, \\
n_i &\to& n_i+\dot \xi^j \bar g_{ij}-\bar N^2 c^2\del_i \phi, \\
h_{ij} &\to&  h_{ij}+2\nabla_{(i} \xi_{j)},
\eea
we find that
\be
\delta_2 S \to \delta_2 S+ \int dt d^3x \sqrt{\bar g}\bar Nc^2 \left[ \frac{\kappa^2}{2}(1-\lambda) \left(\frac{2}{\bar N} \nabla^i(\bar N^2 \nabla_i \phi) \delta K+\frac{c^2}{\bar N^2}(\nabla^i(\bar N^2 \nabla_i \phi))^2\right)-\phi\left(\frac{\dot \rho}{c^2}+\frac{ \nabla_i(\bar N^2 v^i)}{\bar N}\right)\right],
\ee
where we have made use of the energy conservation laws (\ref{con1}) and (\ref{con2}). The action is manifestly invariant under (\ref{shiftn}) to (\ref{shifth}), along with the following shifts in the Stuckelberg fields
\be
\xi^i \to \xi^i-\zeta^i, \qquad \phi \to \phi-f.
\ee
The first Stuckelberg field $\xi^i$ clearly plays no role. Not so the other Stuckelberg field, $\phi$. Its equation of motion is given by
\be
\kappa^2(1-\lambda)\nabla_i\left[\bar N^2\nabla^i \left(\delta
  K+\frac{c^2}{\bar N}\nabla^j(\bar N^2 \nabla_j
  \phi)\right)\right]=\frac{\bar N \dot \rho}{c^2}+\nabla_i(\bar N^2 v^i)+\textrm{non-linear corrections},
\ee
where we have included contributions from terms in the action beyond quadratic order. Now as $\lambda \to 1$, we see that the Stuckelberg field becomes strongly coupled, in direct analogy with the longitudinal scalar degree of freedom in Pauli Fierz gravity. The matter contribution makes this manifest.  Indeed, when matter is present, we can even see the strong coupling of the scalar mode directly from the linearised equations of motion. To see this, consider linearised perturbations that are scalars with respect to the $3D$ diffeomorphisms on spatial slices,
\be
\delta N=n, \qquad \delta N_i=\nabla_i \alpha, \qquad h_{ij}=\sigma \bar g_{ij}+\nabla_i \nabla_j \theta
\ee
It is convenient to make use of the remnant diffeomorphism (\ref{shiftHn}) to gauge away $\theta$. The linearised Hamiltonian constraint (\ref{hamcon}) now yields
\be
\left(\nabla^2+\frac{3}{2} \gamma\right)\sigma=\frac{8 \pi G_N}{c}\rho \label{sol}
\ee
where we have expressed everything in terms of the emergent speed of light (\ref{c}) and Newton constant, $G_N=\kappa^2/32\pi c$. Given the  linearised form of the $N_i$ equation (\ref{Ni}),
\be
\nabla_j \delta \pi^{ij}=8 \pi G_N c v^i
\ee
we make use of the solution (\ref{sol}) and the equation of motion (\ref{eomN}) for $\bar N$, to show that
\be
\bar N (1-\lambda)\left(\nabla^2+\frac{3}{2} \gamma\right)(3\dot
\sigma +2\nabla^2 \alpha)=\frac{16\pi G_N}{c}\left(\bar N \dot \rho+c^2 \nabla_i(\bar N^2 v^i)\right) \label{alpha}
\ee
Now in General Relativity where one has the full set of $4D$ diffeomorphisms, the right hand side of the above equation vanishes automatically by energy-conservation, $\hat \nabla_\mu T^{\mu\nu}=0$, and is therefore consistent with $\lambda \equiv 1$. However, in Ho\v rava gravity, with a reduced set of diffeomorphisms, the reduced version of energy-conservation (\ref{con1}) merely requires $\int d^3x \sqrt{g}\bar N \dot \rho=0$ on this background, and places no constraint on $\nabla_i(\bar N^2 v^i)$. Therefore, by introducing, say,  a non-zero value for $\nabla_i(\bar N^2 v^i)$, the scalar field equation (\ref{alpha}) clearly runs into problems with strong coupling as we approach  the desired infra-red fixed point, $\lambda \to 1$.

Of course, it is important to note that strong coupling will even be present for vacuum fluctuations.  Naively one might expect that we can alleviate the problem by simply absorbing $(1-\lambda)$ into the Stuckelberg field, defining $\hat \phi=\phi(1-\lambda)$. However, the non-linear corrections will generically include terms that schematically go like \mbox{$\kappa^2 (1-\lambda)c^{2m_c}(\del_t)^{m_t}(\nabla)^{m_x}(h_{ij})^{m_{h}}(n_i)^{(5-4m_c-3m_{t}-m_x+3m_\phi)/2}(n)^{m_n}(\phi)^{m_\phi}$}. Upon replacing $\phi$ with $\hat \phi$, such a term contains an overall factor of $(1-\lambda)^{1-m_{\phi}}$, and will diverge for $m_\phi \geq 2$. Of course, this ought to be checked explicitly by introducing the Stuckelberg fields beyond linear order, and computing the higher order action, but this is beyond the scope of the current paper.

Note that unlike the previous case, the strong coupling associated with the Stuckelberg field has nothing to do with detailed balance. It is merely an artifact of the reduced set of diffeomorphisms present in the theory, and occurs even when detailed balance is broken. As one approaches  $\lambda \to 1$, General Relativity is not recovered because the extra degrees of freedom present in the full theory do not all decouple.  On the contrary, one of those degrees of freedom becomes strongly coupled, and one recovers General Relativity with an additional strongly coupled scalar.

\section{Discussion} \label{sec:conclusions}
By considering perturbations about the vacuum we have shown that Ho\v
rava gravity generically suffers from strong coupling problems on all
scales, essentially ruling out the theory as a viable  model of the
Universe. The strong coupling problems come in two different
guises. The first problem is related to the principle of detailed
balance, and can be alleviated by adding terms to the action that
explicitly break this principle. This radically increases the number of parameters one can introduce into the model, and although this would be undesirable from an aesthetic perspective, one could take the view that it would be a small price to pay for a viable model of quantum gravity. Unfortunately breaking detailed balance is not enough, since it does not save us from the second of our strong coupling problems. This is related to the fact that Lorentz invariance is explicitly broken in the UV and one is forced to give up the full set of diffeomorphisms present in General Relativity.  The result is that there are extra degrees of freedom that can still propagate in the infra-red, one of which becomes strongly coupled on all scales as the parameters in the theory approach their desired infra-red fixed point.

Whilst we have explicitly shown these effects for a particular model, we note that they are generic to any model based on Ho\v rava's ideas. Consider first the strong coupling problem associated with detailed balance. Whatever the choice of superpotential, $W[g]$, for detailed balance, the Hamiltonian constraint is given by
\be
-\frac{2}{\kappa^2}K_{ij}G^{ijkl}K_{kl}-\frac{\kappa^2}{8}E^{ij}G_{ijkl}E^{kl}=\rho.
\ee
The vacuum solution ($\rho=0$) is given by $\bar K_{ij}=0$, and so $\bar E^{ij}=0$. Perturbations about the vacuum now yield
\be
-\frac{4}{\kappa^2}\bar K_{ij}\bar G^{ijkl}\delta K_{kl}-\frac{\kappa^2}{4}\bar E^{ij}\bar G_{ijkl}\delta E^{kl}=\rho +\textrm{non-linear corrections}.
\ee
Clearly the left-hand side of the above equation vanishes automatically, which is precisely the first strong coupling issue seen in section~\ref{sec:strongHoravaGravity}.

We now turn our attention to the strong coupling associated with broken diffeomorphism invariance. This is present even without detailed balance, and regardless of how one breaks it. To see this, note that for the theory to have any hope of recovering GR in the infra-red, the quadratic action must take the form $S=S_{GR}+S_{UV}+S_m$, where $S_{UV} \ll S_{GR}$ in the infra-red. Now given the form of the kinetic term in Ho\v rava's model,
\be
S_{UV}= \int dt d^3x \sqrt{\bar g}\bar N\left[ \frac{\kappa^2}{2}(1-\lambda) (\delta K)^2+\textrm{UV corrections coming from the potential}\right].
\ee
Assuming one breaks detailed balance such that the potential is still just a function of the spatial metric and its spatial derivatives, then regardless of its precise form, one will find that the UV corrections above will be invariant under $h_{ij} \to h_{ij}+2\nabla_{(i} \zeta_{j)}$, and as such unaffected by the Stuckelberg fields. The only terms that result in explicit dependence on those fields are $S_m$ and the $(\delta K)^2$ term above. Therefore the Stuckelberg analysis carried out in the previous section can be extrapolated to apply to any breaking of detailed balance, and one recovers the strong coupling problem as $\lambda \to 1$.

We should also comment on our choice of vacuum, since our results clearly depend on this. We believe it is a natural choice since we require the conjugate momenta to vanish on the spatial slices, along with spatial inhomogeneities. This choice admits maximally symmetric spacetimes in $3+1$ dimensions, which are the appropriate vacua in General Relativity. Furthermore, this choice of vacuum, implementing a homogeneous and totally geodesic foliation, is in accord with the Parametrised Post Newtonian (PPN) coordinate system and its basic hypothesis of weak gravity and slowly moving sources. It also contains the Minkowski inertial vacuum for $\gamma=0$. Of course, one could always foliate a maximally symmetric spacetime along surfaces with non-vanishing extrinsic curvature.  It is difficult to see how this would correspond to a better choice of vacuum since the conjugate momenta no longer vanish and we move away from testable regions of GR. In any case, one could always work on temporal scales much larger than the scale set by the extrinsic curvature and reapply our analysis on those scales.  This would presumably set the strong coupling scale to be in the inverse of the extrinsic curvature scale. For example, using a cosmological slicing of de Sitter space would result in strong coupling problems inside the cosmological horizon.

The strong coupling problems guarantee that perturbative General Relativity cannot be reproduced in the infra-red in Ho\v rava gravity. This would seem to disagree with the results of~\cite{Lu:2009em} that recover the Schwarzschild solution when one breaks detailed balance. However, there is no disagreement. The symmetries imposed on the solutions in~\cite{Lu:2009em}  prevent the strongly coupled scalar mode from being excited. Therefore, evidence of this mode may well be absent in classical local tests of general relativity that implement weak and slowly moving sources. Generically, however, the troublesome scalar will be excited. If, for example,  we allowed for time dependence, while keeping spherical symmetry, one would expect this scalar mode to kick in and be responsible for a breaking of Birkhoff's theorem. Indeed the presence of a strong coupled scalar mode in the gravity spectrum casts  serious doubts on the validity of this theorem and signals the probable presence of gravitational radiation from spherical sources.   Furthermore, the linearised version of General Relativity is used to study the effects of gravitational radiation emitted by binary pulsars, and contains excellent agreement with observation~\cite{peri}. In Ho\v rava gravity we have seen that we have no reliable linearised theory to work with due to strong coupling of an extra scalar degree of freedom. Even if it were tractable, it seems unlikely that a non-linear analysis could recover the successes of General Relativity in this instance, since the gravitons will generically couple to the strongly coupled mode through higher order interactions. Our conclusion then is that Ho\v rava gravity in its current form is almost certainly ruled out.

\acknowledgements

We would like to thank Ed Copeland, Nemanja Kaloper and Kazuya Koyama
for useful discussions. AP is funded by a Royal Society University
Research Fellowship, and GN by STFC. CC thanks Elias Kiritsis and Olindo Corradini for early discussions on the subject of Ho\v rava gravity, and for the hospitality shown by Nottingham University during the inception of this work.




\end{document}